# Dynamical Interplay between Coexisting Orders in the Electron-Doped Cuprate Superconductor Nd$_{2-x}$Ce$_x$CuO$_4$

J.P. Hinton,[1,2] J.D. Koralek,[2] G. Yu,[3] E. M. Motoyama,[4] Y.M. Lu,[1] A. Vishwanath,[1] M. Greven,[3] and J. Orenstein[1,2]

[1] Department of Physics, UC Berkeley, Berkeley, CA 94720

[2] Lawrence Berkeley National Lab, Berkeley, CA 94720

[3] School of Physics and Astronomy, University of Minnesota, Minneapolis, MN 55455

[4] Department of Physics, Stanford University, Stanford, CA 94305

We use coherent pump-probe spectroscopy to measure the photoinduced reflectivity $\Delta R$, and complex dielectric function, $\delta\epsilon$, of the electron-doped cuprate superconductor Nd$_{2-x}$Ce$_x$CuO$_{4+\delta}$ at a value of $x$ near optimal doping, as a function of time, temperature, and laser fluence. We observe the onset of a negative $\Delta R$ at $T = 85$ K, above the superconducting transition temperature, $T_c$, of 23 K, whose time dependence exhibits scaling consistent with critical fluctuations. A positive $\Delta R$ onsets at $T_c$ that we associate with superconducting order. We find that the two signals are strongly coupled below $T_c$, in a manner that suggests a repulsive interaction between superconductivity and antiferromagnetic correlations.

Historically, research on high-transition temperature (high-$T_c$), superconductors has pursued two ambitious goals: uncovering their diverse properties, while at the same time identifying those that are essential to superconductivity. The latter path leads naturally to searching for those features that are common to the electron and hole-doped cuprates, as well as iron-pnictide superconductors. Certainly the most discussed of these features, aside from superconductivity itself, are the pseudogap (PG) phenomena. However, in contrast with their prominence in hole-doped cuprates, PG phenomena in the electron-doped cuprates [1] and iron-based superconductors are less well explored. In hole-doped cuprates, transient reflection spectroscopy has proven to be a powerful method for tracking the onsets of both PG and superconducting (SC) order with a single technique [2-6]. Here we report time-resolved reflection (TRR) measurements on the electron-doped superconductor Nd$_{2-x}$Ce$_x$CuO$_{4+\delta}$ (NCCO) in which we observe features strikingly similar those previously associated with the PG in hole-doped cuprates. Moreover, we observe that very low fluence, approximately 2 μJoule cm$^{-2}$ per laser pulse, is sufficient to vaporize the SC condensate in NCCO. We use this effect to measure the temperature dependence of the PG signal below $T_c$ as the strength of SC order is tuned by photoexcitation.

Crystals of NCCO were grown in a traveling-solvent floating-zone furnace in an oxygen atmosphere of 5 bar. To remove the excess oxygen and achieve superconductivity, the crystals were annealed for 10 h in flowing argon at 970°C, followed by 20 h in flowing oxygen at 500°C [7]. The Ce concentration was measured by atomic emission spectrometry. In this paper we focus on samples near optimal doping, for which $x = 0.156$ and $T_c = 23$ K. Our measurements were performed using a mode-locked Ti:Sapphire oscillator generating pulses of 800 nm wavelength light of duration of 150 fs and repetition rate 80 MHz. A pulse picker was employed in order to lower the repetition rate and control average laser heating effects.

Fig. 1a shows the temperature dependence of the maximum amplitude of the transient reflectivity, $\Delta R$, induced by a fluence of 0.5 μJ/cm². We observe two distinct signals of opposite sign, as is observed as well in the hole-doped single-layer cuprate $Pb_{0.55}Bi_{1.5}Sr_{1.6}La_{0.4}CuO_{6+\delta}$ [6] and double-layer $Bi_2Sr_2CaCu_2O_{8+\delta}$ [2,5]. We associate the positive signal that appears at $T_c$ with superconductivity and the negative signal with the PG, as its onset at 85 K lies on an extrapolation of the pseudogap temperature $T^*(x)$ determined from the appearance of gaps in optical conductivity [8] and photoemission [9] spectra.

In attempting to characterize the relationship between the PG and SC responses, it is important to recognize first that the sign of $\Delta R$ does not fully specify the underlying observables, which are the amplitude and phase of the transient reflection coefficient, $\delta r$. The signal measured with conventional pump-probe spectroscopy depends on both through the relation, $\Delta R = 2|r||\delta r| \cos \varphi$, where $\varphi$ is the phase difference between $\delta r$ and the equilibrium coefficient, $r$. The observation that the PG and SC contributions have opposite sign indicates only that the sign of $\cos \varphi$ is reversed for the two components; the relative phase of the two contributions is undetermined.

Coherent transient grating spectroscopy (TGS) provides additional information, in the form of a direct measurement of the phase difference between $\delta r$ and $r$ [10]. In TGS, $\delta r$ is probed by two beams with a phase difference that can be continuously tuned by varying their relative path length. The signals carried by the two probe beams are given by $\Delta R_\pm = 2|r||\delta r| \cos(\varphi \pm \varphi_L)$, where $\varphi_L$ is the path-length controllable optical phase. To determine $\varphi$ we use two photodetectors and measure $\Delta R_+$ and $\Delta R_-$ simultaneously. Fig. 1b shows the output of the two detectors, at a fixed time delay of 1 ps, as a function of $\varphi_L$ for three temperatures spanning $T_c$. The difference in phase between the two cosine waves is equal to twice $\varphi$.

The value of $\varphi$ determined by TGS is plotted as a function of temperature in the inset to Fig. 1a. We observe that $\varphi$ changes by $\pi \pm 0.15$ radian upon crossing from the normal state to the superconducting state. With this knowledge of the relative phase of $\delta r$ and $r$, we can obtain the absolute phase of both $\delta r_{SC}$ and $\delta r_{PG}$, in as much as the phase of $r$ is known from equilibrium reflection spectroscopy and Kramers-Kronig analysis [8]. Finally, the absolute phase of $\delta r$ can be inverted to obtain the phase of the underlying change in the dielectric function for both SC and PG signals, $\delta\epsilon_{SC}$ and $\delta\epsilon_{PG}$, respectively. The parameters thus obtained are plotted as vectors in the complex plane in Fig. 1b. Both real and imaginary parts of $\delta\epsilon_{SC}(1.5\ eV)$ are positive, while for $\delta\epsilon_{PG}(1.5\ eV)$ they are both negative.

Focusing on the phase of the SC response first, it is known from probing at terahertz frequencies [11] that photoexcitation removes spectral weight (SW) from the condensate component of the dielectric function at zero frequency. Our finding that the imaginary part of $\delta\epsilon_{SC}$ is positive indicates that a fraction of the displaced SW reappears at 1.5 eV. This conclusion is consistent with equilibrium optical measurements of hole-doped cuprates such as $Bi_2Sr_2CaCu_2O_{8+\delta}$, which indicate that some of the condensate SW that is lost with increasing $T$ also flows to the frequency scale of the charge transfer gap [12]. Later, we discuss the contrast between the phases of $\delta\epsilon_{SC}$ and $\delta\epsilon_{PG}$.

We turn next to the temperature dependence of the PG and SC signals, beginning with the normal state, where $\Delta R_{SC} = 0$. In Fig. 2a we plot $\Delta R_{PG}(t)$ for several temperatures in the range $T_c < T < T^*$. The time-resolved data show that the increase in amplitude of the transient reflectivity with decreasing $T$ is accompanied by a slowing of response time. In fact, we find that $\Delta R_{PG}(t, T)$ exhibits a strong form of scaling, such that its amplitude and time dependence are described with a single parameter, $\tau_{PG}$. Fig. 2b shows that the data in Fig. 2a collapse to a single curve of the form, $\Delta R_{PG}(t,T) = A\tau_{PG}(T)f(t/\tau_{PG})$, where $A$ is a constant and $f(x) = x\exp(-x)$. The Fig. 2b inset shows that $\tau_{PG}$ grows in proportion to $1/T$ in the normal state, as expected for the characteristic time scale of a system as it approaches a quantum critical point at $T = 0$.

The critical scaling suggests a direct connection between $\Delta R_{PG}(t, T)$ and a fluctuating collective degree of freedom. In NCCO spin fluctuations are a likely candidate, given the experimental evidence [13] and theoretical arguments [14] linking the growth of antiferromagnetic correlations with the pseudogap. However, before we describe how spin fluctuations can reveal themselves in transient reflectivity, we discuss the origin of the scaling form obeyed by $\Delta R_{PG}(t, T)$. For now, we assume that $\Delta R_{PG}(t, T)$ is proportional to the displacement of a collective mode coordinate, $Q$, that obeys damped oscillator dynamics, $(-\omega_0^2 + \omega^2 + i\omega\gamma)Q = F$, where $F$ is a generalized force acting on $Q$. The single parameter scaling form shown in Fig. 2b follows from Eq. 1 at critical damping, $\gamma = 2\omega_0$, in which case the time domain response to a δ-function force takes the form, $Q(t) \propto It \exp(-\gamma t)$, where $I \equiv \int F dt$. Thus we associate the time scale, $\tau_{PG}$, with the inverse damping rate of the fluctuations of $Q$.

As it is reasonable in the case of NCCO to associate the divergent time scale $\tau_{PG}$ with antiferromagnetic fluctuations [13], the question arises as to how the spin degrees of freedom might manifest themselves in an optical pump-probe measurement. A possible answer lies in the close connection between photoinduced reflectivity and the Raman interaction, as was emphasized recently in the context of nonlinear phonon coupling [15]. The Raman interaction for spins on a square lattice has the form [16],

$$H_R = C \sum_i (\boldsymbol{E} \cdot \boldsymbol{n}_i)(\boldsymbol{E} \cdot \boldsymbol{n}_j) \boldsymbol{S}_i \cdot \boldsymbol{S}_j, \qquad (1)$$

where $\boldsymbol{E}$ is the electric field of the light, $\boldsymbol{n}_i$ are the nearest neighbor unit vectors, and $C$ is a coupling constant. This term in the Hamiltonian leads to inelastic scattering of photons accompanied by a pair of opposite spin flips, a process known as two-magnon Raman scattering. If we associate the collective coordinate $Q$ with $\langle \boldsymbol{S}_i \cdot \boldsymbol{S}_j \rangle$, the spin-photon interaction energy is proportional to $E^2 Q$, equivalent to a generalized force acting on $\langle \boldsymbol{S}_i \cdot \boldsymbol{S}_j \rangle$ that is also proportional to $E^2$. As a result of this interaction the *envelope of the pump pulse* acts as an impulse that displaces $\langle S_i \cdot S_j \rangle$ from its value in equilibrium. Another way to envision the same physics is to recognize that Eq. 1 has the form of an exchange Hamiltonian, with the pulse envelope function $E^2$ playing the role of an exchange energy. Thus the pump pulse induces a transient change in the exchange coupling, generating a force which acts to reduce the local antiferromagnetic spin correlation. Moreover, the same Raman interaction accounts for the sensitivity of the probe beam to transient changes in $\langle S_i \cdot S_j \rangle$ that are induced by the pump [17]. In the context of the Hubbard model, the SW of transitions that create double occupancy will decrease with decreasing $\langle S_i \cdot S_j \rangle$, leading to a negative sign for the imaginary part of $\delta\epsilon_{PG}(1.5\ eV)$, consistent with our result depicted in Fig. 1b.

Next we examine the interplay of the PG and SC signals as a function of $T$ and the laser fluence, $\Phi$. Figs. 3a-c show plots of $\Delta R(t)/R$ at several temperatures spanning $T_c$ for pulse fluencies of 2, 4, and 8 μJoules/cm², respectively. Figs. 3d-f depict the same data set in color scale plots of $\Delta R/R$ in the temperature-time plane. In recording these data the pulse repetition rate was varied in inverse proportion to the energy per pulse, such that the average optical power delivered to the sample remained the same. Thus the average temperature rise, estimated at less than 1 K based on the observed shifts in $T_c$, is the same for all the data shown in Fig. 3. Below $T_c$, and at the lowest fluence, the positive SC signal dominates the response and the PG signal is not clearly discernible. As $\Phi$ is increased, the relative strength of the two signals reverses, with $\Delta R/R$ approaching a single component PG response at high fluence.

To determine the relationship between PG and SC responses, we resolve $\Delta R(t)$ into its components, as quantified by the coefficients $A_{SC}$ and $A_{PG}$ in the expression, $\Delta R(t) = A_{PG}\Delta R_{PG}(t) + A_{SC}\Delta R_{SC}(t)$. In performing this decomposition, we assume that $\Delta R_{PG}(t,T)$ obeys the same scaling form as in the normal state. A further assumption is that the rise and decay of $\Delta R_{SC}(t)$ can be described by exponential functions, that is, $\Delta R_{SC}(t) \propto \left(1 - e^{-t/\tau_r}\right)e^{-t/\tau_d}$. Given these forms for the PG and SC components, we then vary the parameters $\tau_d$, $\tau_{PG}$, $A_{PG}$, and $A_{SC}$ to achieve the best fit to $\Delta R(t,T)$. The resulting values of $A_{PG}$ and $A_{SC}$ are largely insensitive to $\tau_r$, which was set to 1.4 ps for the analysis presented here. The dashed lines in Figs. 3a-c illustrate the high quality of the fits obtained by the superposition of PG and SC responses described above (an example of the decomposition is shown in the Fig. 3c inset).

In Figs. 4a and 4b we plot $A_{PG}(T,\Phi)$ and $A_{SC}(T,\Phi)$ as determined by the fitting procedure described above. Fig. 4a shows $A_{PG}$ and $A_{SC}$ as functions of $\Phi$, for a representative temperature above $T_c$ (26 K) and one below $T_c$ (5 K). While above $T_c$, $A_{PG}$ is linear in laser fluence, both components of $\Delta R$ are nonlinear functions of $\Phi$ in the superconducting state. The SC signal plateaus at a rather low saturation fluence, $\Phi_s \sim 1.5$ $\mu$Joule-cm$^{-2}$, a value that is much smaller than found in the higher-$T_c$ materials BSCCO and YBCO, but is consistent with the scaling $\Phi_s \propto T_c^2$ recently reported in hole-doped cuprates and pnictide superconductors [18]. This saturation phenomenon is generally associated with photoinduced vaporization of the SC condensate. In contrast, $A_{PG}$ at 5 K grows superlinearly in the same fluence regime where $A_{SC}$ is sublinear. The clear implication is that with increasing fluence the PG correlations strengthen as SC order weakens, suggestive of a repulsive interaction between two order parameters.

Evidence for repulsive interaction is seen as well in Fig. 4b, which is a plot of the fluence-normalized weighting factors, $A_{PG}/\Phi$ and $A_{SC}/\Phi$, as functions of $T$, for the same three values of $\Phi$ as in Fig. 3. Above $T_c$, $A_{PG}(T)/\Phi$ is independent of fluence, indicating that $\Delta R_{PG}$ is linear in $\Phi$ throughout the normal state. However, at 23 K the normalized amplitudes for PG and SC diverge, demonstrating that nonlinearity appears abruptly at $T_c$. Based our hypothesis relating $\Delta R_{PG}$ to $\langle \mathbf{S_i} \cdot \mathbf{S_j} \rangle$, we would tend to associate the suppression of the PG signal that begins at $T_c$ with weakening of spin correlations in the presence of superconductivity. On the other hand, these data bear a striking resemblance to the $T$ dependence of recently reported fluctuating [19] and static [20] charge density wave order in YBa$_2$Cu$_3$O$_{7-\delta}$, which are also suppressed by the appearance of superconductivity. It is possible that both spin and charge order, with varying amplitudes and correlation timescales, contribute to the PG phenomenology in these systems.

To conclude, we have used time-resolved reflectivity and transient grating spectroscopy to measure the photoinduced reflectivity $\Delta R$, and complex dielectric function, $\delta\epsilon$, as a function of time delay, temperature, and laser fluence in Nd$_{2-x}$Ce$_x$CuO$_4$. We observe the onset of a $\Delta R$ signal, associated with opening of the pseudogap, that exhibits a form of time-temperature scaling consistent with critical fluctuations. To interpret the observation of scaling, we speculate that $\Delta R$ probes local spin correlations through the same spin-photon coupling that leads to two-magnon Raman scattering. If this suggestion is correct, measurement of $\Delta R(t,T)$ can be used to measure spin correlations in the time domain, which would be particularly valuable in critically damped regimes where Raman scattering observes only a broad background. We note that strongly damped dynamics have been predicted for $\langle \boldsymbol{S_i} \cdot \boldsymbol{S_j} \rangle$ in connection with the theory of Raman scattering near a quantum critical point [21]. At $T_c$ we observe a second component of $\Delta R$ that is clearly associated with superconductivity. Using a coherent pump-probe technique we determine that the two components of $\Delta R$ arise from changes in spectral at 1.5 eV that are opposite sign. Finally we observe that the SC and PG components of the transient reflectivity are strongly coupled below $T_c$, consistent with theory that suggests a repulsive interaction between superconductivity and antiferromagnetic correlations in electron-doped cuprates [22].

**Acknowledgments**

Optical spectrsoscopy research is supported by the Director, Office of Science, Office of Basic Energy Sciences, Materials Sciences and Engineering Division, of the U.S. Department of Energy under Contract No. DE-AC02-05CH11231. The crystal growth and characterization work was supported by NSF Grant DMR-1006617 and by a seed grant through the NSF MRSEC program.

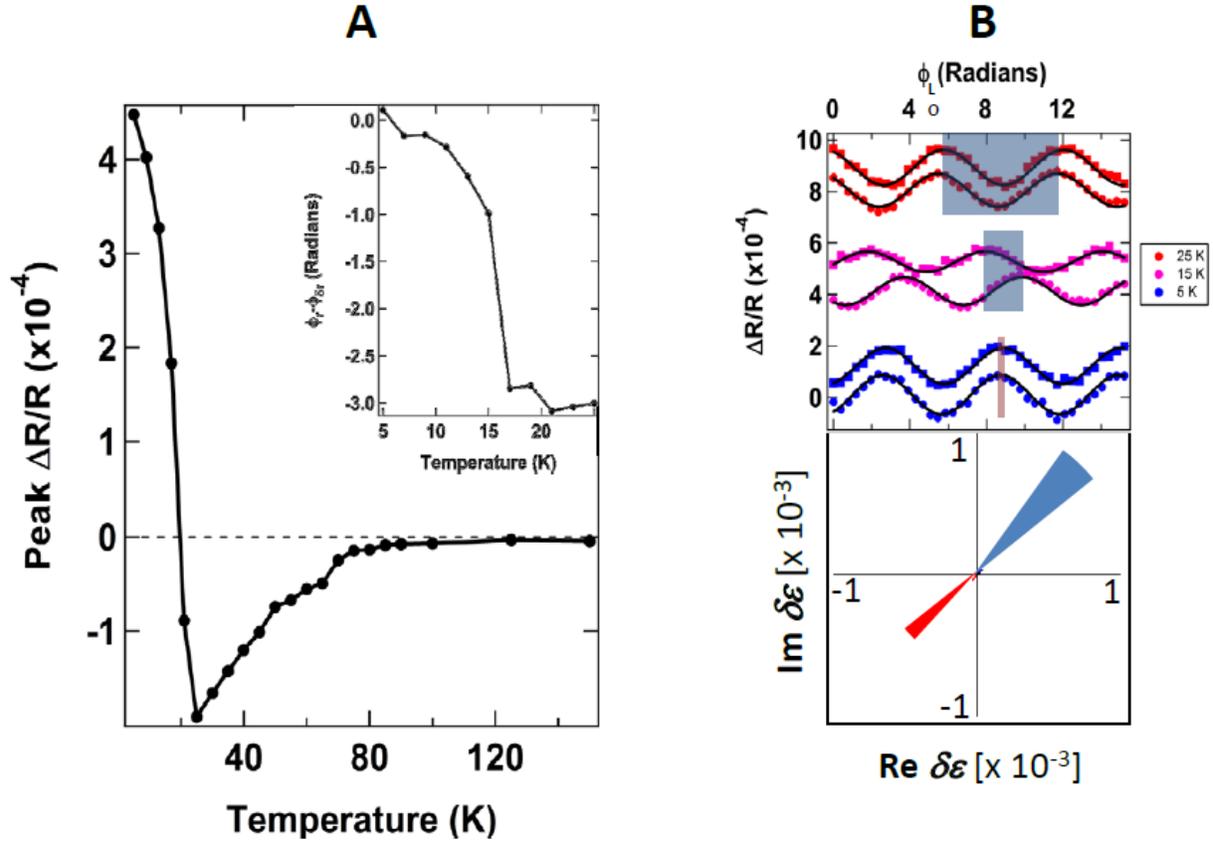

**Figure 1. (A)** The amplitude of the fractional change in reflectivity, as a function of temperature measured at a pump laser fluence of 0.6 µJ-cm$^{-2}$, showing the onset of a negative signal in the normal state at 85 K and a positive signal near the superconducting transition temperature of 23 K. **(B)** Measurements using coherent transient grating spectroscopy. Upper panel: dependence of the transient grating signal on the relative phase of the two phase-conjugate probe beams. The phase shift between the two beams, shown as a shaded region, is equal to twice the phase difference between the equilibrium reflection coefficient, $r$, and the photoinduced change, $\delta r$. The relative phase of $r$ and $\delta r$ is plotted as a function of temperature in the inset of Fig. 1a. Lower panel: The change in the dielectric function at 1.5 eV that corresponds to the transient reflectivity is plotted in the complex plane for the PG signal (red) and SC signal (blue). The triangular shapes indicate the experimental uncertainty in the determining the absolute phase.

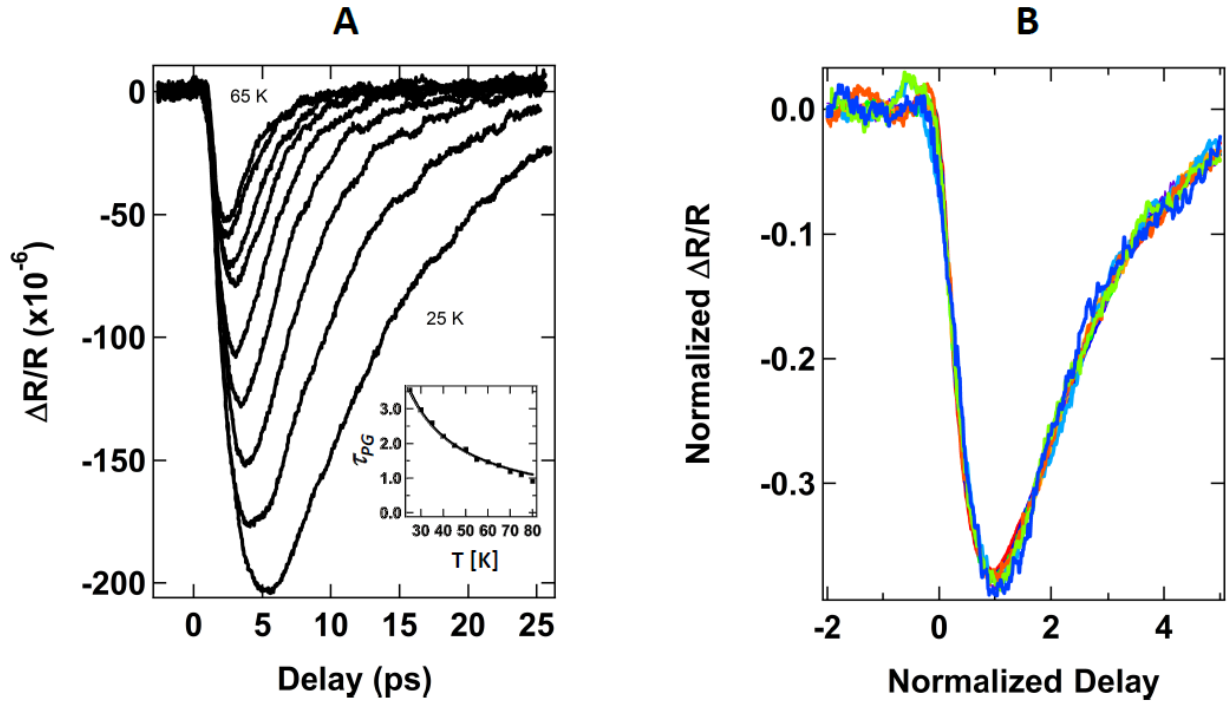

**Figure 2.** **(A)** Fractional change in reflectivity in the normal state, shown for temperatures from 25 and 65 K (with intervals of 5 K). **(B)** Single parameter data collapse to scaling form described in the text. The decay time of each curve is plotted vs. temperature in the inset to **(A)**, together with a solid line that indicates a $1/T$ dependence on temperature.

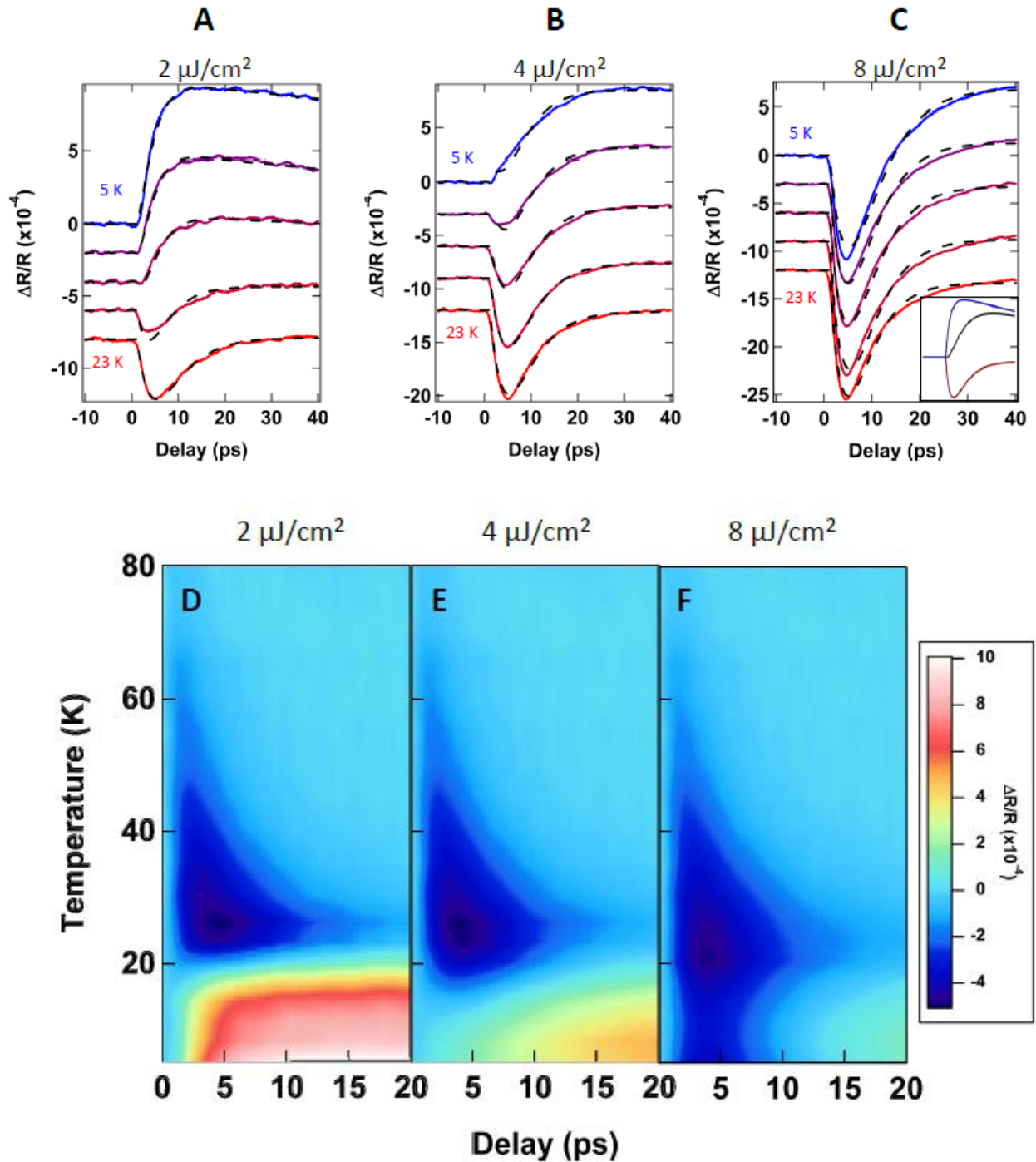

**Figure 3.** The time delay and temperature dependence for laser fluence of **(A)** 2 µJ/cm², **(B)** 4 µJ/cm², and **(C)** 8 µJ/cm². Fits using procedure described in text are shown by the dashed lines. The inset to **(C)** shows the decomposition of the signal measured at 5 K and 4 µJ/cm² into PG (red) and SC (blue) components. **(D)**, **(E)**, and **(F)** are the same data sets plotted as false color images.

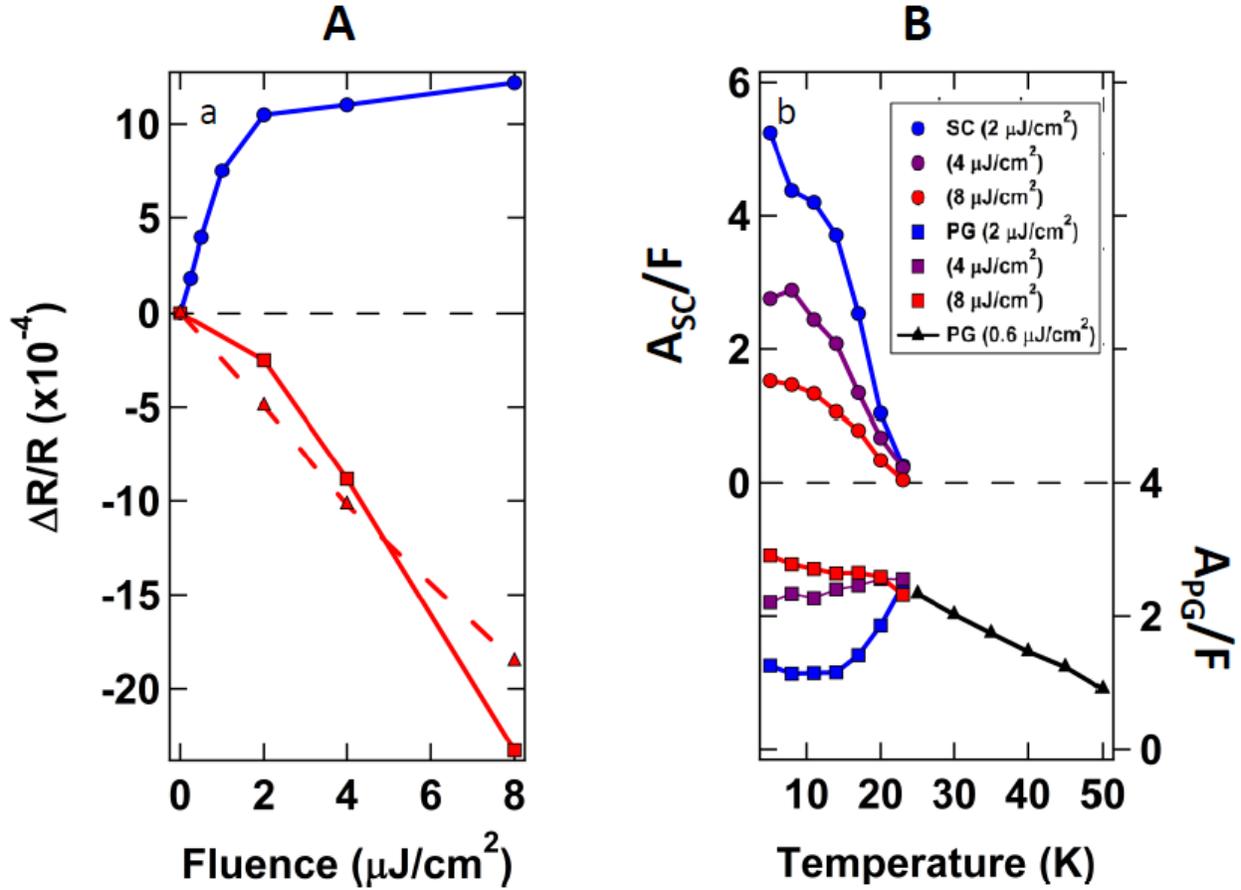

**Figure 4. (A)** The PG (red squares) and SC (blue circles) components of the transient reflectivity as a function of laser fluence at 5 K, compared with the normal state signal (red triangles) as measured at 23 K. In the normal state the transient reflectivity is linear in fluence. Below $T_c$ the SC component saturates while the PG component exhibits a superlinear dependence on fluence. **(B)** The fluence-normalized SC (left axis) and PG (right axis) components as a function of temperature. At the lowest fluence, 2 µJoule/cm$^2$, the PG component is suppressed at the onset of superconductivity. The extent of this suppression becomes smaller as increased laser fluence causes a weakening of superconducting order. Lines in **(A)** and **(B)** are guides to the eye.